\documentclass{PoS}

\title{DVCS off deuteron and twist three contributions}

\ShortTitle{DVCS off deuteron and twist $3$}

\author{\speaker{I.~V.~Anikin}\\
        Bogoliubov Laboratory of Theoretical Physics, JINR,
             141980 Dubna, Russia\\
        E-mail: \email{anikin@theor.jinr.ru}}

\author{R.~S.~Pasechnik\\
        Department of Physics and Astronomy, Uppsala
             University,  SE-751 20 Uppsala, Sweden\\
        E-mail: \email{roman.pasechnik@fysast.uu.se}}

\author{B.~Pire\\
        CPHT, \'Ecole Polytechnique, CNRS,
             91128 Palaiseau, France\\
        E-mail: \email{pire@cpht.polytechnique.fr}}

\author{O.~V.~Teryaev\\
        Bogoliubov Laboratory of Theoretical Physics, JINR,
             141980 Dubna, Russia\\
        E-mail: \email{teryaev@theor.jinr.ru}}

\abstract{
We study the deeply virtual Compton scattering off a spin-one particle, which is exemplified 
by the case of
coherent scattering on a deuteron target. We discuss the role of twist three contributions
for restoring the QED gauge invariance of the amplitude corresponding to this process. We
consider both kinematical and dynamical sources of twist three generalized parton
distributions. 
}

\FullConference{35th International Conference of High Energy Physics - ICHEP2010,\\
        July 22-28, 2010\\
        Paris France}

\begin{document}


\section{Introduction.}

Deeply virtual Compton scattering (DVCS) on the deuteron target has recently attracted
much attention  from the experimental point of view
\cite{ Mazouz,  HERMES-deuteron}.
One of the main reasons of this interest is the fact
that the DVCS process gives information about a new type of
parton distributions, called generalized parton distributions (GPDs)
which allow to extract much information about the quark and gluon structure of hadrons,
particularly its spin structure \cite{TL}, 
and to allow a femtophotography of nuclei \cite{femto}.

From the theoretical point of view, the leading twist-2 GPDs for the
deuteron were defined in \cite{Berger} and the DVCS amplitude on the
deuteron was discussed at leading order in  \cite{Cano}. However,
the leading twist-2 accuracy for the DVCS amplitude is not enough
for the study of such processes with significant transverse momenta,
because of the QED gauge invariance breaking of the DVCS amplitude
in leading twist-2 order in the Bjorken limit and non zero
transverse final momenta. This problem was resolved in \cite{APT}
for a (pseudo)scalar target  (pion, $He^4$), where it was
demonstrated that one can restore the gauge invariance of the  DVCS
amplitude by taking into account the twist-3 contributions, related
to the matrix elements of quark-gluon operators. Besides, the
relevant additional terms
 provide the leading contribution to some polarization observables.
Then, the same ideas  were used and generalized for the nucleon target
\cite{Pol-tw3}.
We here follow  the approach, presented in \cite{APT},
to make a comprehensive analysis of the twist three contributions to the
amplitude of the DVCS off deuteron (or an arbitrary spin-one target).



\begin{figure}[t]
\centerline{\includegraphics[width=0.3\textwidth]{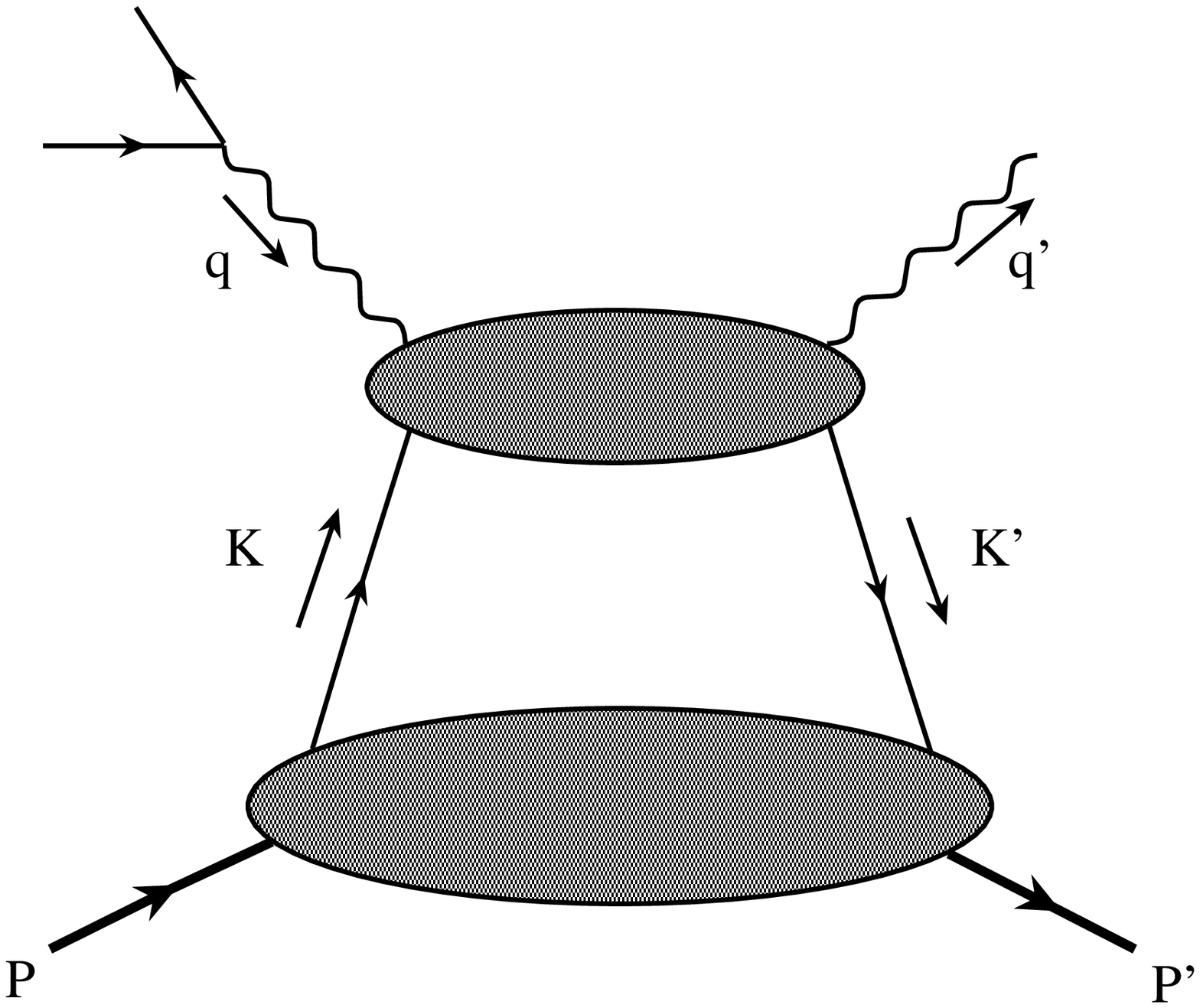}
\hspace{1.cm}\includegraphics[width=0.3\textwidth]{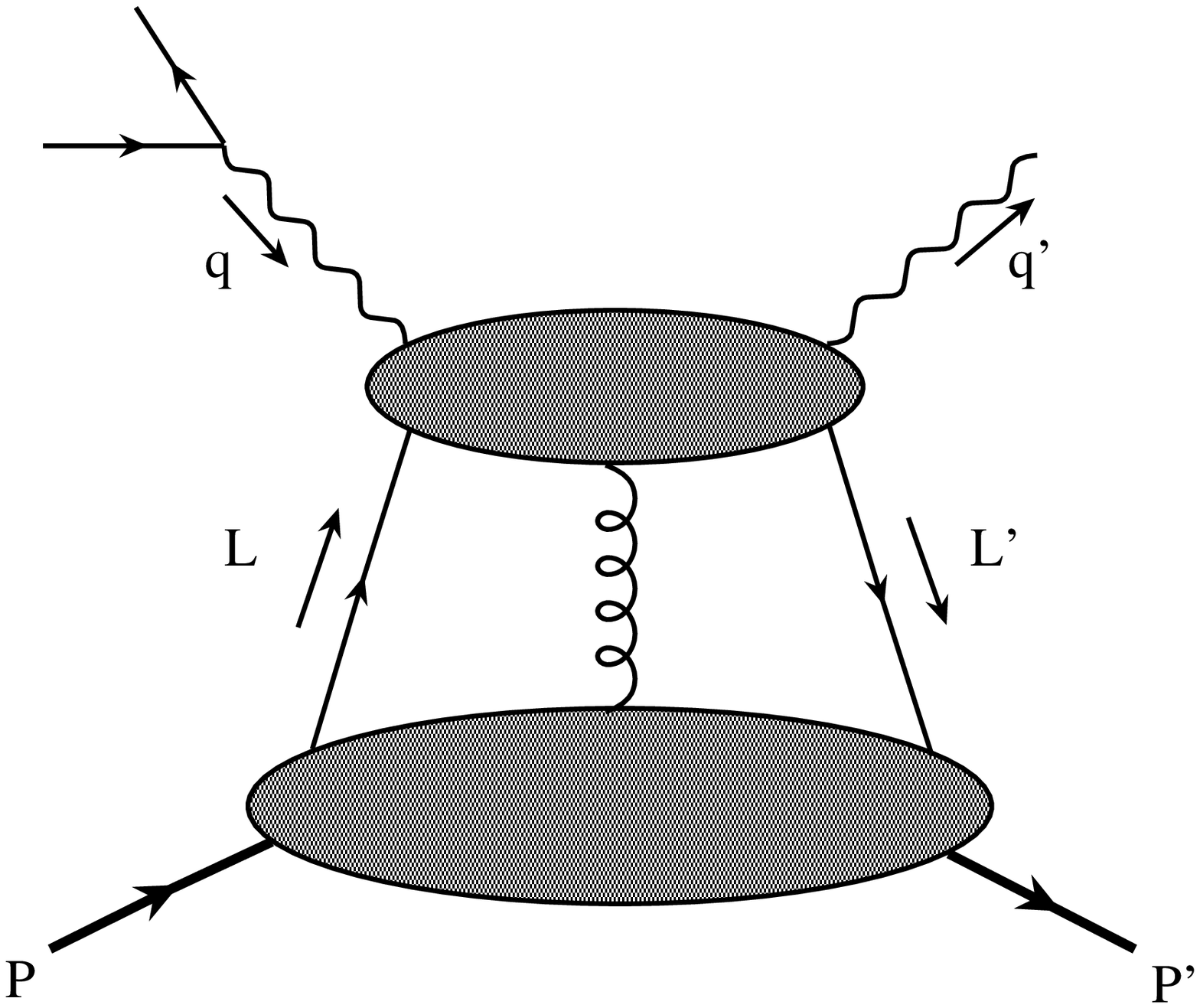}}
  \caption{The Feynman diagrams corresponding to  deeply virtual Compton scattering.
Notations: $P\equiv p_1,\quad P^{\prime}\equiv p_2,\quad K\equiv
k-\Delta/2\approx xP-\Delta/2, \quad K^{\prime}\equiv
k+\Delta/2\approx xP+\Delta/2, \quad L\equiv k_1-\Delta/2\approx
x_1P-\Delta/2, \quad L^\prime\equiv k_2+\Delta/2\approx
x_2P+\Delta/2$. Here, $k$ and $k_i$ correspond to the loop momenta
in the diagrams.} \label{Fig1}
\end{figure}

Let us start with the discussion of the kinematics and approximations
which we use in this paper. The process we consider is
$$\gamma^*(q) + D(p_1) \to \gamma(q^\prime) + D(p_2),$$
with $q^2=-Q^2$ large, while $q^{\prime\, 2}=0$. At the Born level,
the Feynman diagrams corresponding to this process are depicted in
Fig.~\ref{Fig1}. We introduce the ``plus'' and ``minus'' vectors as
$n^\star=\Lambda(1,\, 0, \, 0,\, 1)\, , \quad n=1/(2\Lambda)(1,\, 0,
\, 0,\, -1)\, , \quad n^\star\cdot n =1$. We consider the DVCS
amplitude up to the twist three accuracy, discarding the
contributions associated with the twist four and higher. The hadron
relative and transfer momenta can be written as
\begin{eqnarray}
\label{kin} &&P=\frac{p_1+p_2}{2}=n^\star+\frac{\bar M^2}{2}n\,
\approx \, n^\star \,,\quad \Delta=p_2-p_1 = -2\xi P + 2\xi\bar M^2
n +\Delta^T \approx -2\xi P +\Delta^T,
\nonumber\\
&&\xi=\frac{(p_1-p_2)^+}{(p_2+p_1)^+}, \quad P\cdot \Delta=0, \quad
\Delta^2=t=\Delta_T^2-4\xi^2 \bar M^2 \approx 0 .
\end{eqnarray}


\section{Parameterization of the vector and  axial-vector matrix elements}


We now introduce the parameterization of all relevant matrix elements up to
the twist three accuracy.
The parameterization of the twist-$2$ vector correlators is standard  and
can be found in \cite{Berger}, for which
we will use the shorthand notation:
\begin{eqnarray}
\label{VparTw2}
\langle p_2,\lambda_2|
\left[ \bar\psi(0)\gamma_{\mu} \psi(z) \right]^{tw-2}
| p_1,\lambda_1 \rangle
\stackrel{{\cal F}_1}{=}
P_{\mu}\, H^V_{1,..,5}(e^{*}_2, e_1; x,\xi,t).
\end{eqnarray}
We now come to the discussion of the twist-$3$ operator matrix elements and their parametrizations.
We parametrize the vector quark correlator as \footnote{the symbol $\stackrel{{\cal F}_1}{=}$ denotes the Fourier transformation with the measure:
$dx \,\exp\{ -i(xP-\Delta/2)\cdot z\}$, while $\stackrel{{\cal F}_2}{=}$ corresponds to the integration with
$dx_1 dx_2 \exp\{ -i(x_1 P-\Delta/2)\cdot z_1 - i(x_2-x_1) P\cdot z_2\}$.}
\begin{eqnarray}
\label{VparTw3}
&&\langle p_2,\lambda_2|
\left[ \bar\psi(0)\gamma_{\mu} \psi(z) \right]^{tw-3}
| p_1 ,\lambda_1 \rangle
\stackrel{{\cal F}_1}{=}
\Delta^T_{\mu}\, G^{V}_{1,..,5}(e^{*}_2, e_1; x,\xi) +
e^{*\,T}_{2\,\mu} (e_1\cdot P)\, G^{V}_6(x,\xi) +
\nonumber\\
&&e^{T}_{1\,\mu} (e^*_2\cdot P)\,  G^{V}_7(x,\xi) +
M^2\, e^{*\, T}_{2\, \mu} (e_1\cdot n)\, G^{V}_8(x,\xi)+
M^2\, e^{T}_{1\, \mu} (e^*_2\cdot n)\, G^{V}_9(x,\xi) .
\end{eqnarray}
In the forward limit, where $\Delta=0$, one has
$P_\mu \Rightarrow p_\mu , \, e^{*}_{2 \, \mu} \Rightarrow e^*_\mu ,
\, e_{1\, \mu} \Rightarrow e_\mu $.
Therefore, in this limit, the parameterizations (\ref{VparTw2}) and (\ref{VparTw3}) reduce to
the parameterizations with $H^V_1(x,0),\, H^V_5(x,0)$ and $G^V_8(x,0), \, G^V_9(x,0)$.
The deuteron, as a spin-one particle, has its polarization degrees of freedom
described by the spin density matrix:
$e^*_\mu e_\nu = P_{\mu\nu}/3 + S_{\mu\nu} +
i/(2M)\,\varepsilon_{\mu\nu\alpha\beta} S_\alpha p_\beta$,
where $P_{\mu\nu}$ is the well-known unpolarized projector, the vector $S_{\mu}$ represents the vectorial
polarization and the tensor $S_{\mu\nu}$ -- the tensorial one.
With this, one can see that the twist-$2$ corresponds to either the unpolarized or
the tensorial-polarized deuteron, while the twist-$3$ describes the tensorial or vectorial
polarization.

As in \cite{APT}, we introduce the matrix elements with
a transverse derivative (detailed consideration of such matrix elements
can be found in \cite{AT, AIPSW}). The parameterization of the quark-antiquark correlator
with the transverse derivative is written as  the following nine terms:
\begin{eqnarray}
\label{dTparTw3V}
&&\langle p_2,\lambda_2|
\left[ \bar\psi(0)\gamma_{\mu} i\partial^T_\rho \psi(z) \right]^{tw-3}
| p_1 ,\lambda_1 \rangle
\stackrel{{\cal F}_1}{=}
P_\mu \biggl\{ \Delta^T_{\rho} \, b^T_{1,..,5}(e^{*}_2, e_1; x,\xi) +
e^{*\,T}_{2\,\rho} (e_1\cdot P)\,  b^T_6(x,\xi) +
\biggr.
\nonumber\\
\biggl.
&&e^{T}_{1\,\rho} (e^*_2\cdot P)\, b^T_7(x,\xi) +
M^2\, e^{*\, T}_{2\, \rho} (e_1\cdot n)\, b^T_8(x,\xi)+
M^2\, e^{T}_{1\, \rho} (e^{*}_2\cdot n)\, b^T_9(x,\xi) \biggr\} .
\end{eqnarray}
In a similar way, we  parametrize the quark-antiquark-gluon correlator of
 genuine twist 3, replacing in the r.h.s. of Eq. (\ref{dTparTw3V})
 $b^T_i(x,\xi)$ by $B_i(x_1, x_2,\xi)$.



The twist-$2$ axial-vector correlator is also standard one and
can be parametrized as in \cite{Berger} by
\begin{eqnarray}
\label{AparTw2}
&&\langle p_2,\lambda_2|
\left[ \bar\psi(0)\gamma_{\mu}\gamma_5 \psi(z) \right]^{tw-2}
| p_1 ,\lambda_1 \rangle
\stackrel{{\cal F}_1}{=}
-i\, e^{*}_{2\,\alpha} \, {\cal A}^{(i),\,L}_{\alpha\beta,\,\mu}
(n^\star,n,\Delta_T) e_{1\, \beta}\,H^{A}_i(x,\xi,t).
\end{eqnarray}
In the forward limit, the twist-$2$ axial-vector correlator
corresponds to the case where the deuteron has the (longitudinal) vectorial polarization.
For the twist-3 correlators, we have, using the Schouten identity to determine
the Lorentz independent structures,
\begin{eqnarray}
\label{AparTw3}
&& i \,\langle p_2,\lambda_2|
\left[ \bar\psi(0)\gamma_{\mu}\gamma_5 \psi(z) \right]^{tw-3}
| p_1 ,\lambda_1 \rangle
\stackrel{{\cal F}_1}{=}
\varepsilon_{\mu n P e_1^T}(e^*_2\cdot P)\, G^A_1(x,\xi) +
\varepsilon_{\mu n P e^{*\,T}_2} (e_1\cdot P)\, G^A_2(x,\xi)
\nonumber\\
&+&M^2\, \varepsilon_{\mu n P e^T_1} (e^*_2\cdot n)\, G^A_3(x,\xi) +
M^2\, \varepsilon_{\mu n P e^{*\, T}_2} (e_1\cdot n)\, G^A_4(x,\xi) +
\frac{1}{M^2}\,\varepsilon_{\mu\Delta_T P e^*_2} (e_1\cdot P)\, G^A_5(x,\xi) +
\\
&&
\varepsilon_{\mu\Delta_T P e^*_2} e_1\cdot n\, G^A_6(x,\xi) +
 \varepsilon_{\mu\Delta_T P e_1} e^*_2\cdot n\, G^A_7(x,\xi) +
\varepsilon_{\mu\Delta_T n e^{*}_2} e_1\cdot P\, G^A_8(x,\xi) +
M^2\, \varepsilon_{\mu \Delta_T n e_1} e^*_2\cdot n\, G^A_9(x,\xi) . \nonumber
\end{eqnarray}
Again, in the forward limit, these twist-$3$ correlators are related
to the tensorial or (transverse) vectorial polarizations of
deuteron. The matrix element of the twist-$3$ operator associated
with the quark-antiquark operator containing a transverse derivative
reads
\begin{eqnarray}
\label{dTparTw3A}
&&i \,\langle p_2,\lambda_2|
\left[ \bar\psi(0)\gamma_{\mu}\gamma_5 i\partial^T_\rho \psi(z) \right]^{tw-3}
| p_1 ,\lambda_1 \rangle
\stackrel{{\cal F}_1}{=}
P_\mu\biggl\{
\varepsilon_{\rho n P e_1^T}(e^*_2\cdot P)\, d^T_1(x,\xi) +
\varepsilon_{\rho n P e^{*\,T}_2} (e_1\cdot P)\, d^T_2(x,\xi)
\biggr.
\nonumber\\
\biggl.
&+&M^2\, \varepsilon_{\rho n P e^T_1} (e^*_2\cdot n)\, d^T_3(x,\xi) +
M^2\, \varepsilon_{\rho n P e^{*\, T}_2} (e_1\cdot n)\, d^T_4(x,\xi) +
\frac{1}{M^2}\,\varepsilon_{\rho\Delta_T P e^*_2} (e_1\cdot P)\, d^T_5(x,\xi)
\biggr.
\nonumber\\
\biggl.
&+&
\varepsilon_{\rho\Delta_T P e^*_2} e_1\cdot n\, d^T_6(x,\xi) +
 \varepsilon_{\rho\Delta_T P e_1} e^*_2\cdot n\, d^T_7(x,\xi) +
\varepsilon_{\rho\Delta_T n e^{*}_2} e_1\cdot P\, d^T_8(x,\xi)
\biggr.
\nonumber\\
\biggl.
&+&
M^2\, \varepsilon_{\rho\Delta_T n e_1} e^*_2\cdot n\, d^T_9(x,\xi)
 \biggr\} .
\end{eqnarray}
From (\ref{dTparTw3A}), it is not difficult to
parameterize the three-particle correlator with the
genuine twist-$3$.

\section{Gauge invariant amplitude of DVCS on the deuteron target}

Taking into account both kinematical and dynamical twist-$3$ contributions and
using the QCD equations of motion  relating the twist $2$ and $3$ (see \cite{APT}),
the gauge invariant expression of the DVCS amplitude
 takes the form:
\begin{eqnarray}
\label{amp}
&&T_{\mu\nu}^{(\lambda_1,\, \lambda_2)} =
\frac{1}{2P\cdot \bar Q}\int dx
\frac{1}{x-\xi+i\epsilon}
\Biggl({\cal T}^{(1)}_{\mu\nu}+{\cal T}^{(2)}_{\mu\nu} +{\cal T}^{(3)}_{\mu\nu}
+{\cal T}^{(4)}_{\mu\nu}\Biggr)^{(\lambda_1,\, \lambda_2)}
+ O(\Delta^2_T;\, \bar M^2) \,
+``crossed"
\nonumber\\
\end{eqnarray}
where $\bar Q=(q+q')/2$ and the structure amplitudes ${\cal
T}^{(k)}_{\mu\nu}$ read
\begin{eqnarray}
&&{\cal T}^{(1)}_{\mu\nu} =
H^V_{1,..,5}(x; e_1, e^*_2)
\Biggl(
2\xi P_{\mu}P_{\nu}  +
P_{\mu}\bar Q_{\nu} + P_{\nu}\bar Q_{\mu} -
g_{\mu\nu}(P\cdot \bar Q) + \frac{1}{2}P_{\mu}\Delta_{\nu}^{T} -
\frac{1}{2}P_{\nu}\Delta_{\mu}^{T}
\Biggr) +
\nonumber\\
&&G^V_{1,..,5}(x; e_1, e^*_2)
\Biggl(
\xi P_{\nu}\Delta_{\mu}^T + 3\xi P_{\mu}\Delta_{\nu}^T
+ \Delta_{\mu}^{T}\bar Q_{\nu} + \Delta_{\nu}^{T} \bar Q_{\mu}
\Biggr) +
\nonumber\\
&&
\Biggl( M^2(e_1\cdot n) (e^*_2\cdot n) G^A_9(x) -
\frac{(e^*_2\cdot P)(e_1\cdot P)}{M^2}G^A_5(x)-
(e^*_2\cdot P)(e_1\cdot n)G^A_6(x)-
\Biggr.
\nonumber\\
\Biggl.
&&
(e_1\cdot P)(e^*_2\cdot n)\left( G^A_7(x)-G^A_8(x)\right)
\Biggr)
\Biggl(
3\xi P_{\mu}\Delta_{\nu}^{T} -
\xi P_{\nu}\Delta_{\mu}^{T} -
\Delta_{\mu}^{T}\bar Q_{\nu} + \Delta_{\nu}^{T}\bar Q_{\mu}
\Biggr) ,
\nonumber\\
&& {\cal T}^{(2)}_{\mu\nu} =
\Biggl((e_1\cdot P)G^V_6(x) +
M^2 (e_1\cdot n) G^V_{8}(x)\Biggr)
\Biggl(
\xi P_{\nu} e^{*\,T}_{2\, \mu} + 3\xi P_{\mu}e^{*\,T}_{2\, \nu}
+ e^{*\,T}_{2\, \mu}\bar Q_{\nu} + e^{*\,T}_{2\, \nu} \bar Q_{\mu}
\Biggr) +
\nonumber\\
&&
\Biggl((e_1\cdot P)G^A_2(x) +
M^2(e_1\cdot n)G^A_4(x)
\Biggr)
\Biggl(
3\xi P_{\mu}e^{*\,T}_{2\, \nu} -
\xi P_{\nu}e^{*\,T}_{2\, \mu} -
e^{*\,T}_{2\, \mu} \bar Q_{\nu} + e^{*\,T}_{2\, \nu} \bar Q_{\mu}
\Biggr) ,
\nonumber\\
&& {\cal T}^{(3)}_{\mu\nu} =
\Biggl(
(e^{*}_2\cdot P)G^V_7(x)+
M^2 (e^*_2\cdot n) G^V_{9}(x)
\Biggr)
\Biggl(
\xi P_{\nu} e^{T}_{1\, \mu} + 3\xi P_{\mu}e^{T}_{1\, \nu}
+ e^{T}_{1\, \mu}\bar Q_{\nu} + e^{T}_{1\, \nu} \bar Q_{\mu}
\Biggr) +
\nonumber\\
&&
\Biggl(
(e^{*}_2\cdot P)G^A_1(x)+
M^2(e^*_2\cdot n)G^A_3(x)
\Biggr)
\Biggl(
3\xi P_{\mu}e^{T}_{1\, \nu} -
\xi P_{\nu}e^{T}_{1\, \mu} -
e^{T}_{1\, \mu} \bar Q_{\nu} + e^{T}_{1\, \nu} \bar Q_{\mu}
\Biggr)
\nonumber\\
&& {\cal T}^{(4)}_{\mu\nu} =
\varepsilon_{\mu\nu P n}
\Biggl(
\varepsilon_{n P e^{*\,T}_2 e^{T}_1}\, H^A_1(x,\xi) +
\frac{1}{M^2}\, \varepsilon_{n P \Delta^T e^{*\,T}_2} (e_1\cdot P)\, H^A_2(x,\xi) +
\Biggr.
\nonumber\\
\Biggl.
&&\frac{1}{M^2}\,\varepsilon_{n P \Delta^T e^T_1} (e^*_2\cdot P)\, H^A_3(x,\xi) +
\varepsilon_{n P \Delta^T e^{*\,T}_2} (e_1\cdot n)\, H^A_4(x,\xi)
\Biggr)
\nonumber
\end{eqnarray}



In conclusion, we have taken into account both the kinematical and dynamical twist-$3$ contributions
in order to derive the gauge invariant amplitude of the deeply virtual Compton scattering
off a spin-one particle (deuteron).


The authors would like to thank A.V.~Efremov, D.~Ivanov, L.~Szymanowski, S.~Wallon
for useful discussions and comments.
 This work is supported in part by the  contract
18853PJ, the RFBR (grants 09-02-01149, 09-02-00732) and the Carl
Trygger Foundation.


\begin{thebibliography}{99}




\bibitem{Mazouz}

 M.~Mazouz {\it et al.}  [Jefferson Lab Hall A Collaboration],
  Phys.\ Rev.\ Lett.\  {\bf 99}, 242501 (2007).



\bibitem{HERMES-deuteron}
A.~Airapetian  [The HERMES Collaboration],
 Nucl.\ Phys.\  B {\bf 829}, 1 (2010)
and   B {\bf 842}, 265 (2011).

 \bibitem{TL}
 S.~K.~Taneja and S.~Liuti,
  arXiv:1008.1706 [hep-ph].

  \bibitem{femto}
   J.~P.~Ralston and B.~Pire,
  Phys.\ Rev.\  D {\bf 66}, 111501 (2002).

\bibitem{Berger}
  E.~R.~Berger, F.~Cano, M.~Diehl and B.~Pire,
  Phys.\ Rev.\ Lett.\  {\bf 87}, 142302 (2001).





\bibitem{Cano}
  F.~Cano and B.~Pire,
  Eur.\ Phys.\ J.\  A {\bf 19}, 423 (2004);
 A.~Kirchner and D.~Mueller,
  Eur.\ Phys.\ J.\  C {\bf 32}, 347 (2003)].


\bibitem{APT}
 I.~V.~Anikin, B.~Pire and O.~V.~Teryaev,
 Phys.\ Rev.\  D {\bf 62}, 071501 (2000).



\bibitem{Pol-tw3}

  M.~Penttinen  {\it et.al.},
  Phys.\ Lett.\  B {\bf 491}, 96 (2000);
 A.~V.~Belitsky {\it et.al.},
  Nucl.\ Phys.\  B {\bf 593}, 289 (2001);
 A.~V.~Belitsky and D.~Mueller,
  Nucl.\ Phys.\  B {\bf 589}, 611 (2000).
  M.~Vanderhaeghen,
  Eur.\ Phys.\ J.\  A {\bf 8}, 455 (2000);
N.~Kivel {\it et.al.},
  Phys.\ Lett.\  B {\bf 497}, 73 (2001);
  A.~V.~Radyushkin and C.~Weiss,
  Phys.\ Lett.\  B {\bf 493}, 332 (2000).






\bibitem{AT}

  I.~V.~Anikin and O.~V.~Teryaev,
  Phys.\ Lett.\  B {\bf 509}, 95 (2001).




\bibitem{AIPSW}
  I.~V.~Anikin {\it et.al.},
 Phys.\ Lett.\  B {\bf 682}, 413 (2010)
and
  Nucl.\ Phys.\  B {\bf 828}, 1 (2010).

\bibitem{Efr84} A.V. Efremov, O.V. Teryaev, Yad. Phys. {\bf 39} (1984) 1517.



\end{thebibliography}
\end{document}